\pdfoutput=1
\documentclass{JHEP3}
\usepackage{subfigure}
\usepackage{graphicx}
\usepackage{mcite}
\usepackage{amsmath}
\usepackage{amssymb}
\usepackage{amsfonts}
\usepackage{mathrsfs}
\usepackage{multirow}
\usepackage{ulem}
\newcommand{\inputTikZ}[1]{\includegraphics{#1}}

\newcommand{\ev}[1]{\ensuremath{\left\langle #1 %
                     \right\rangle}} 
\newcommand{\tr}{{\rm tr}} 

\newcommand{\Rep}[1]{\ensuremath{\underline{\mathbf{#1}}}}

\newcommand{\SU}[1]{\ensuremath{\mathrm{SU}(#1)}}
\newcommand{\U}[1]{\ensuremath{\mathrm{U}(#1)}}
\newcommand{\SO}[1]{\ensuremath{\mathrm{SO}(#1)}}

\newcommand{\I}{\ensuremath{\mathrm{i}}}
\newcommand{\diag}{\ensuremath{\mathrm{diag}}}

\newcommand{\Eqref}[1]{Eq.~\eqref{#1}}
\newcommand{\Figref}[1]{Fig.~\ref{#1}}
\newcommand{\Tabref}[1]{Tab.~\ref{#1}}
\newcommand{\Tabrefs}[3]{Tabs.~\ref{#1}, \ref{#2} and \ref{#3}}
\newcommand{\Secref}[1]{Sec.~\ref{#1}}
\newcommand{\Appref}[1]{App.~\ref{#1}}

\renewcommand{\mathbb}[1]{{\mathbf{#1}}}

\renewcommand\arraystretch{1.3}

\newcommand{\MPI}{Max--Planck--Institut f\"{u}r Kernphysik, Postfach 10 39 80,
69029 Heidelberg, Germany}
\newcommand{\IPPP}{Institute for Particle Physics Phenomenology, University of
Durham, DH1 3LE, Durham, United Kingdom}

\title{\mathversion{bold}Flavored Orbifold GUT -- an SO(10)$\times S_4$ Model}
\author{Adisorn Adulpravitchai\\\MPI\\Email: \email{a.adulpravitchai@mpi-hd.mpg.de}}
\author{Michael A.~Schmidt\\\IPPP\\Email: \email{m.a.schmidt@durham.ac.uk}}
\preprint{IPPP/10/02; DCPT/10/04}
\keywords{Beyond Standard Model, Orbifold GUT, Neutrino Physics,
  Discrete Flavor Symmetry, Tribimaximal Mixing}

\abstract{Orbifold grand unified theories (GUTs) solve several problems in GUT model
building. Therefore, it is intriguing to investigate similar constructions in
the flavor context. In this paper, we propose that a flavor symmetry might
emerge
due to  orbifold compactification of one orbifold and broken by  boundary
conditions of
another orbifold. The combination of the orbifold parities in gauge and flavor
space determines the zero
modes. We demonstrate the construction in a supersymmetric (SUSY) SO(10)$\times
S_4$ orbifold GUT model, which predicts the tribimaximal mixing at leading order in the lepton sector 
as well as the Cabibbo angle in the quark sector.}

\begin{document}

\section{Introduction}
The leptonic mixing matrix is compatible with the tribimaximal mixing
matrix~\cite{Harrison:2002er} and hints towards a flavor symmetry. Especially,
non-abelian discrete flavor symmetries have been studied. However, the origin of
a potential flavor
symmetry is unclear. The embedding in a continuous gauge symmetry seems
disfavored because the explanation of the flavor structure requires large
representations of flavon fields which cannot couple directly to the three
generations of the SM fermions~\cite{Adulpravitchai:2009kd,*Berger:2009tt}. In
the context of string theory, string selection rules may lead to a discrete
flavor symmetry~\cite{Kobayashi:2004ya,Kobayashi:2006wq,Ko:2007dz}. Moreover,
in magnetized extra dimensional models, a discrete flavor symmetry may arise
from the localization behavior of zero
modes~\cite{Abe:2009vi,*Abe:2010ii}. In addition, a discrete symmetry might
be a remnant of an orbifold
compactification~\cite{Dixon:1986qv,Watari:2002fd,*Watari:2002tf,
Altarelli:2006kg,Kobayashi:2006wq,Adulpravitchai:2009id}, which we
consider in this study. 
Orbifolds, even several, are used in heterotic string theory model building and
can
therefore be consistently embedded in a UV complete theory.
An orbifold compactification of a GUT can lead to its
breaking and nicely solve e.g.~the doublet-triplet splitting
problem~\cite{Kawamura:2000ev}. Recently, a flavor symmetry originating from an
orbifold
compactification has been studied in a GUT context~\cite{Burrows:2009pi}.
A gauge symmetry is broken in an orbifold construction by the non-trivial
transformation of the gauge fields under the orbifold
parities~\cite{Scherk:1978ta,*Scherk:1979zr} or Wilson
lines~\cite{Ibanez:1986tp}. Similarly, it is possible to break
a discrete flavor symmetry by a non-trivial transformation of the bulk
fermions~\cite{Haba:2006dz} as well as by Wilson lines~\cite{Seidl:2008yf}.
Alternatively, the orbifold
compactification can generate the alignment of vacuum expectation values (VEVs)
of flavons~\cite{Kobayashi:2008ih} transmitting the flavor symmetry breaking
into the fermion mass matrices.\\
In this paper, we study the combination of the origin of a flavor symmetry as
well as its breaking by the VEV alignment of flavons from an orbifold.  We
assume two orbifolds, where the flavor symmetry originates from the special
geometry of one orbifold and it is broken on another orbifold. We demonstrate it
with a simple model in the context of a SUSY SO(10) orbifold
GUT with $S_4$ flavor
symmetry~\cite{Pakvasa:1978tx,*Yamanaka:1981pa,*Brown:1984mq,*Brown:1984dk,
*Lee:1994qx,*Ma:2005pd,*Hagedorn:2006ug,*Cai:2006mf,*Caravaglios:2006aq,
*Zhang:2006fv,*Koide:2007sr,*Parida:2008pu,*Lam:2008sh,*Bazzocchi:2008ej,
*Ishimori:2008fi,*Bazzocchi:2009da,*Altarelli:2009gn,*Ishimori:2009ns,
*Grimus:2009pg,*Ding:2009iy,*Meloni:2009cz,*Morisi:2010rk,Dutta:2009bj}. The
model predicts the tribimaximal mixing at leading order in the lepton
sector and the Cabibbo angle in the quark sector by implementing the Gatto-Sartori-Tonin relation~\cite{Gatto:1968ss,*Oakes:1969vm}. Note
that this does not depend on SUSY. We assume that the orbifold parities act on
gauge, SUSY as well as
flavor space. Hence, the zero modes are determined by the overall parity.

\section{Flavor Symmetry from Orbifolding\label{sec:origin}}
\begin{figure}
 \centering
 \subfigure{\inputTikZ{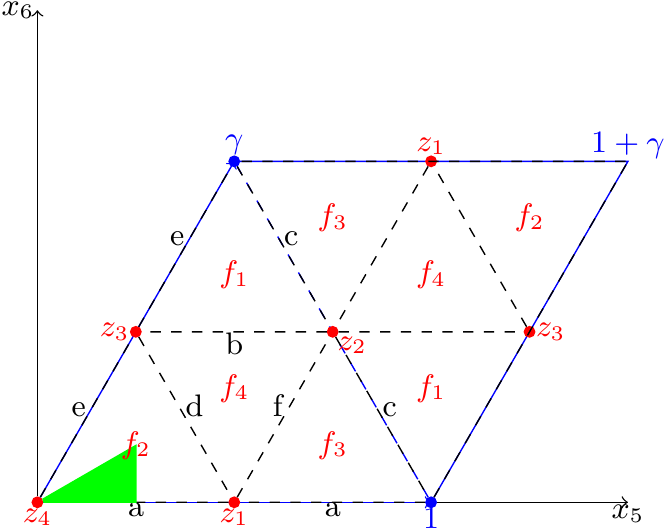}}
\subfigure{\inputTikZ{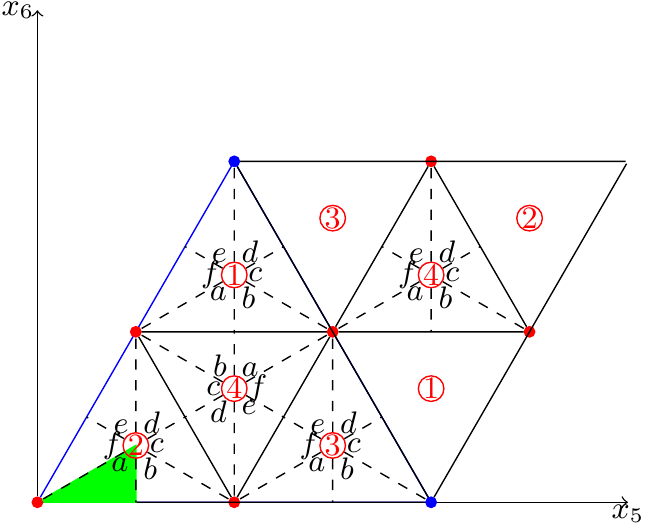}}
 \caption{The orbifold $T^2/\mathbb{Z}_2$ with its four fixed points
$(z_1,\,z_2,\,z_3,\,z_4)$. The
    symmetry  of interchanging the fixed points forms the discrete group $S_4$.
$S_4$ is a symmetry of the whole orbifold, where the edges
$(a,\,b,\,c,\,d,\,e,\,f)$ form a six dimensional representation and the
triangles building up the faces a 24 dimensional representation.}
 \label{fig:OrbifoldSO10andS4}
\end{figure}

We study the
$T^2/\mathbb{Z}_2$
orbifold (see \Figref{fig:OrbifoldSO10andS4}) with radii $R=R_5=R_6$, which we
choose as $2\pi R=1$ for simplicity. The discussion of the flavor
symmetry does not change for $2\pi R\neq 1$. It is defined by
\begin{align}
T_1: z \rightarrow& z+1, &
T_2: z \rightarrow& z+ \gamma, &
Z: z \rightarrow& -z.
\end{align}
where $z=x_5+\I\, x_6$ and $\gamma=e^{\I \pi/3}$. The shape of this orbifold is
a regular tetrahedron. This choice of equal radii and $\gamma=e^{\I\pi/3}$ is
motivated by the possibility to stabilize the shape by Casimir energy as
discussed in~\cite{Ponton:2001hq,*Buchmuller:2009er}.
It has been shown in~\cite{Altarelli:2006kg} that the breaking of Poincar\'e
symmetry from 6d to 4d through compactification on the orbifold leads to a
remnant $S_4$ flavor symmetry. Concretely, the orbifold has four fixed points,
\mbox{$(z_1,z_2,z_3,z_4)=(1/2,\,(1+\gamma)/2,\,\gamma/2,\,0)$}, which
are permuted by two translation operations $S_i$, the rotation $T_R$, and two
parity operations $P^{(\prime)}$
\begin{align}
S_1:&z \rightarrow z + 1/2,\;\; &
S_2:&z \rightarrow z + \gamma/2, \;\; &
T_R:&z \rightarrow \gamma^2 z, &
P:& z \rightarrow z^{*}, \;\; &
P^\prime:& z \rightarrow -z^{*}\; .
\end{align}
One can also write these operations explicitly in terms of the interchange of
the fixed points, $S_1[(14)(23)]$, $S_2[(12)(34)]$, $T_R[(123)(4)]$,
$P[(23)(1)(4)]$ and $P^\prime[(23)(1)(4)]$.
From these elements we can define two generators of $S_4$ as $S = S_2 P$ and $T
= T_R$ satisfying the generator relation,
$S^4=T^3=(ST^2)^2=1$. The localization of a brane field defines its
representation of $S_4$.
Concretely, the generators $S$, $T$ can be represented by the matrices,
\begin{align}
 S= \begin{pmatrix} 0 & 0 & 1 & 0 \\ 1 & 0 & 0 & 0 \\ 0 & 0 & 0 & 1 \\ 0 & 1 & 0
& 0 \end{pmatrix}, & \;\;\;\;
 T= \begin{pmatrix} 0 & 0 & 1 & 0 \\ 1 & 0 & 0 & 0 \\ 0 & 1 & 0 & 0 \\ 0 & 0 & 0
& 1 \end{pmatrix},
\end{align}
acting on the brane field $\psi(x_\mu)=(\psi_1,\psi_2,\psi_3,\psi_4)^T$, where
$\psi_{i}=\psi(x_\mu,z_i)$ is a field localized at fixed point $z_i$. We
denote this basis as localization basis. The characters of $S$ and $T$ show that
the four dimensional representation generated by $\left<S,\,T\right>$ can be
decomposed in
$3_1\oplus1_1$. The explicit unitary transformation is
\begin{equation}
 S \rightarrow U^{\dagger} S U = \begin{pmatrix}  S^{fl}_3 &  \\   & 1
\end{pmatrix},\quad
 T \rightarrow U^{\dagger} T U = \begin{pmatrix}   T^{fl}_3 &  \\  & 1
\end{pmatrix}\quad
\end{equation}
with $U=\begin{pmatrix}(\alpha_{ij}) & (\beta_i)\\\end{pmatrix}$ and
\begin{equation}
(\alpha_{ij})=\begin{pmatrix}
 -\frac{1}{2 \sqrt{3}} & \frac{1}{\sqrt{3}} & \frac{1}{\sqrt{3}}\\
 -\frac{1}{2 \sqrt{3}} & \frac{\omega}{\sqrt{3}} & \frac{\omega^2}{\sqrt{3}} \\
 -\frac{1}{2 \sqrt{3}} & \frac{\omega^2}{\sqrt{3}} & \frac{\omega}{\sqrt{3}} \\
 \frac{\sqrt{3}}{2} & 0 & 0 \end{pmatrix},\quad\quad
(\beta_i)=\begin{pmatrix}
\frac{1}{2} \\
 \frac{1}{2}\\
\frac{1}{2} \\
\frac{1}{2} \end{pmatrix},
\end{equation}
where $\omega=e^{2\pi\I/3}$ and $S^{fl}_3,T^{fl}_3$ are the three dimensional
generators of $S_4$ acting on a triplet
$3_1$ in flavor basis.
The transformation of a field $\psi(x)$ is accordingly related to
the flavor basis $\psi^{fl}(x_\mu)=U^\dagger\psi(x_\mu)$ as well as the
Clebsch-Gordan coefficients. The first three components of $\psi^{fl}$ form a
triplet $3_1$ and the last one a singlet $1_1$.
It is possible to remove one of the representations from the
  low-energy spectrum by adding a bulk field transforming as $3_1$ ($1_1$) and oppositely charged to the
brane field, such that they acquire a Dirac mass term.

The representations $1_2$ and $3_2$ are analogously obtained by using the
freedom to change the phase of each brane field in a symmetry
transformation. Concretely, by replacing $S$ with $-S$, we obtain a
representation of $S_4$ which decomposes as
$\left<-S,\,T\right>=3_2\oplus 1_2$. Similarly, the $2$ of $S_4$ can be obtained
from the four dimensional representation  $\left<S:=T_R^2\,
P\,T_R,\,T:=T_R^2\right>=2\oplus1_1\oplus1_1$.
Analogously, the edges are exchanged by $S[(ab)(cf)(de)]$ and $T[(ace)(bdf)]$.
 $S$ and $T$ form a  six
dimensional representation  which can be expressed in  matrix form by
\begin{align}
S&=\left(
\begin{array}{cccccc}
 0 & 1 & 0 & 0 & 0 & 0 \\
 1 & 0 & 0 & 0 & 0 & 0 \\
 0 & 0 & 0 & 0 & 0 & 1 \\
 0 & 0 & 0 & 0 & 1 & 0 \\
 0 & 0 & 0 & 1 & 0 & 0 \\
 0 & 0 & 1 & 0 & 0 & 0
\end{array}
\right)\;,&
T&= \left(
\begin{array}{cccccc}
 0 & 0 & 0 & 0 & 1 & 0 \\
 0 & 0 & 0 & 0 & 0 & 1 \\
 1 & 0 & 0 & 0 & 0 & 0 \\
 0 & 1 & 0 & 0 & 0 & 0 \\
 0 & 0 & 1 & 0 & 0 & 0 \\
 0 & 0 & 0 & 1 & 0 & 0
\end{array}
\right)\;.
\end{align}
It can be decomposed in 
\begin{equation}
6=2\oplus2\oplus1_1\oplus1_2
\end{equation}
and the explicit unitary transformation which transforms $S$ and $T$ from the
localization basis \mbox{$(S,T)\to(U^\dagger S U, U^\dagger T U)$} into flavor
basis is given by
\begin{equation}
U=\left(
\begin{array}{cccccc}
 0 & \frac{\omega ^2}{\sqrt{3}} & \frac{\omega ^2}{\sqrt{3}} & 0 &
\frac{1}{\sqrt{6}} & -\frac{1}{\sqrt{6}} \\
 \frac{\omega ^2}{\sqrt{3}} & 0 & 0 & \frac{\omega ^2}{\sqrt{3}} &
\frac{1}{\sqrt{6}} & \frac{1}{\sqrt{6}} \\
 0 & \frac{1}{\sqrt{3}} & \frac{\omega }{\sqrt{3}} & 0 & \frac{1}{\sqrt{6}} &
-\frac{1}{\sqrt{6}} \\
 \frac{\omega }{\sqrt{3}} & 0 & 0 & \frac{1}{\sqrt{3}} & \frac{1}{\sqrt{6}} &
\frac{1}{\sqrt{6}} \\
 0 & \frac{\omega }{\sqrt{3}} & \frac{1}{\sqrt{3}} & 0 & \frac{1}{\sqrt{6}} &
-\frac{1}{\sqrt{6}} \\
 \frac{1}{\sqrt{3}} & 0 & 0 & \frac{\omega }{\sqrt{3}} & \frac{1}{\sqrt{6}} &
\frac{1}{\sqrt{6}}
\end{array}
\right)\;.
\end{equation}
Indeed, the symmetry generated by $S:z\to z^*+\gamma/2$ and
$T$ is a symmetry of the whole orbifold, because $S$ and $T$ fulfill the $S_4$
generator relations
\begin{subequations}
\begin{align}
S^4 : z&\stackrel{S}{\longrightarrow} z^*+\gamma/2 \stackrel{S}{\longrightarrow}
z+\gamma^*/2+\gamma/2
\stackrel{S^2}{\longrightarrow}z^*+\gamma^*/2\to z\;,\\
T^3: z&\stackrel{T}{\longrightarrow}\gamma^2 z
\stackrel{T^2}{\longrightarrow} \gamma^6 z =z\;,\\
(ST^2)^2: z&\stackrel{ST^2}{\longrightarrow} \gamma^2 z^* +\gamma/2
 \stackrel{ST^2}{\longrightarrow} \gamma^2(\gamma^2 z^*
+\gamma/2)^*+\gamma/2=z\;.
\end{align}
\end{subequations}
The orbits under the action of $S_4$, i.e. the equivalence classes with respect
to
the group action, can be represented by the points within the green (gray) triangle.
Hence, the fundamental domain is reduced to the green (gray) triangle.
The 6d Poincar\'e symmetry is broken to 4d Poincar\'e symmetry and the
discrete symmetry $S_4$.
The generators can be expressed as $S:[(1243)]\otimes[(ab)(cf)(ed)]$ and
$T:[(123)]\otimes[(ace)(bdf)]$. Hence, the 24 dimensional representation is
constructed
by the tensor product of the four dimensional representation of the vertices
and the six dimensional representation of the edges. It can be decomposed into
irreducible representations as follows
\begin{equation}
\begin{split}
4\otimes6
&=(1_1\oplus3_1)\otimes(1_1\oplus1_2\oplus2\oplus2)\\
&=1_1\oplus3_1\oplus1_2\oplus3_2\oplus2\oplus3_1\oplus3_2\oplus2\oplus3_1\oplus3
_2\; .
\end{split}
\end{equation}
We note that for the remaining sections we will work in the flavor basis with
irreducible
representations only and omit $^{fl}$ for simplicity.

\section{Symmetry Breaking by Boundary Conditions}
\begin{figure}
 \centering
 \inputTikZ{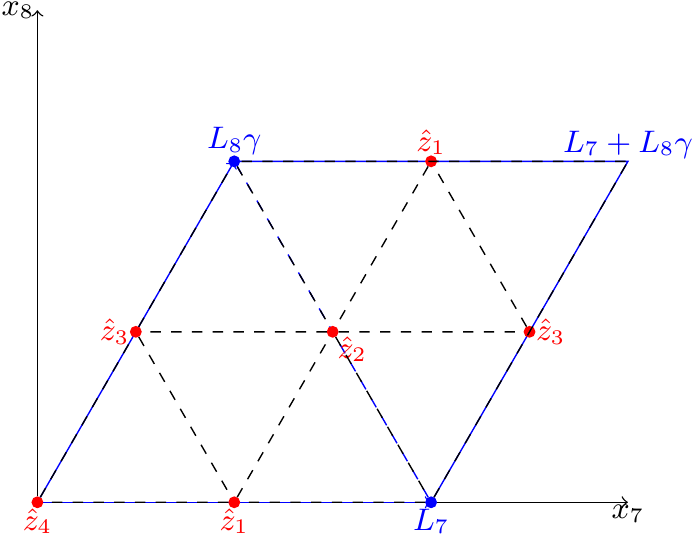}
 \caption{The gauge symmetry SO(10) as well as the flavor symmetry $S_4$ is
broken by boundary conditions on the orbifold $T^2/(\mathbb{Z}_2^I \times
\mathbb{Z}_2^{PS}\times\mathbb{Z}_2^{GG})$.
 \label{fig:OrbifoldBreaking}}
\end{figure}

In order to demonstrate how the flavor structure can be obtained and broken
appropriately from an orbifold, we implement it in an 8d SUSY SO(10) model with a
breaking of gauge symmetry, which nicely leads to a splitting of the
doublet and triplet components in the $\Rep{10}$-plet similar
to the 6d model in~\cite{Asaka:2002nd,*Hall:2001xr}. 
As the different boundary
conditions lift the degeneracy of the fixed points, the flavor symmetry has to
emerge from a different orbifold than the one where it is broken. (We note that
in the first orbifold we can impose only one boundary condition without lifting
the degeneracy of the fixed points, however, this is not enough for breaking the
gauge symmetry, therefore, the second orbifold is needed.) For simplicity, we
assume that the orbifold on which the symmetry is broken is also
$T^2/\mathbb{Z}_2$ with two additional boundary conditions at $\hat{z}_1$,
$\hat{z}_3$, i.e. $T^2/(\mathbb{Z}_2^I \times
\mathbb{Z}_2^{PS}\times\mathbb{Z}_2^{GG})$ (see \Figref{fig:OrbifoldBreaking})
and then we assume that the gauge fields are bulk fields of the two orbifolds,
while all other bulk fields of the second orbifold are brane fields of the first
orbifold.
The $N=1$ SUSY in 8d leads to $N=4$ SUSY in
4d~\cite{ArkaniHamed:2001tb}. Its gauge vector multiplet can be decomposed in
terms of one vector multiplet and three chiral multiplets of the unbroken $N=1$
SUSY in 4d. In order to break $N=4$ SUSY down to $N=1$ SUSY, we use the orbifold
parity of the first orbifold combined with one orbifold parity from the second
orbifold, such that only the vector muliplet of $N=1$ SUSY remains as zero mode.

The resulting GUT scale $M_{GUT}$ is determined by the
compactification scale. We neglect anomaly
cancellation~\cite{Hebecker:2001jb,Asaka:2002my} for the purpose of this study.
The gauge fields transform under the orbifold parities as
\begin{align}
P_0 V(x_\mu,-z,\hat{z})P_0^{-1} =& \eta_0 V(x_\mu,z,\hat{z})\;, \nonumber \\
P_IV(x_\mu,z,-\hat{z})P_I^{-1} =& \eta_I V(x_\mu,z,\hat{z})\;, \nonumber \\
P_{PS}V(x_\mu,z,-\hat{z}+\hat{z}_1)P_{PS}^{-1} =&
\eta_{PS} V(x_\mu,z,\hat{z}+\hat{z}_1)\;,\nonumber \\
P_{GG}V(x_\mu,z,-\hat{z}+\hat{z}_3)P_{GG}^{-1} =&
\eta_{GG} V(x_\mu,z,\hat{z}+\hat{z}_3)\;,
\end{align}
with $\hat{z}=x_7+\I x_8$ where the matrices,
\begin{align}
P_0&=P_I=\mathbb{1}\;,\nonumber\\
P_{PS}&=\diag(-1,-1,-1,1,1)\otimes\sigma^0\;,\nonumber\\
P_{GG}&=\diag(1,1,1,1,1)\otimes\sigma^2\;,
\end{align}
 act on the gauge space with $\sigma^0$ being the $2\times 2$ unity matrix and
$\sigma^2$ one of the Pauli matrices . The parities are chosen as
$\eta_0=\eta_I=\eta_{PS}=\eta_{GG}=+1$.
The first two parities (corresponding to fixed point $z_4$ and $\hat{z}_4$)
are used to break $N=4$ SUSY to $N=1$
SUSY and the remaining two parities are used
to break the gauge symmetry~\cite{Mirabelli:1997aj,Asaka:2001eh}. The parity
assignment of the different components of the bulk fields is given in
\Tabrefs{tab:GaugeField}{tab:bulk10}{tab:bulk16}~\cite{Asaka:2001eh,Asaka:2002my}.
\begin{table}[htb]
\begin{center}
\renewcommand{\arraystretch}{1.5}
\begin{tabular}{|c|c|c||ccc|}
 \hline
 &  &  &
 \multicolumn{3}{ c|}{($V_\mu$,~$\lambda_1$)}
 \\
 \cline{4-6}
 $G_{SM'}$ &
 $G_{GG}$  &
 $G_{PS}$ &
 $Z_2^I$ &
 $Z_2^{GG}$ &
 $Z_2^{PS}$
 \\
 \hline
  \itshape \bfseries
  $( \Rep{8}, \Rep{1}, 0)_0$ & $\Rep{24}_0$ &  $(\Rep{15},\Rep{1},\Rep{1})$
   & $+$ & $+$ & $+$
 \\
 $ ( \Rep{3}, \Rep{2},$-$5)_0$ & $\Rep{24}_0$ & $(\Rep{6},\Rep{2},\Rep{2})$
   & $+$ & $+$ & $-$
 \\
  $( \bar{\Rep{3}}, \Rep{2}, 5)_0$ & $\Rep{24}_0$ & $(\Rep{6},\Rep{2},\Rep{2})$
   & + & + & $-$
 \\
  \itshape \bfseries
  $( \Rep{1}, \Rep{3}, 0)_0$ & $\Rep{24}_0$ & $(\Rep{1},\Rep{3},\Rep{1})$
   & + & + & +
 \\
  \itshape \bfseries
  $( \Rep{1}, \Rep{1}, 0)_0$ & $\Rep{24}_0 $ & $(\Rep{1},\Rep{1},\Rep{3})$
   & + & + & +
 \\
 $ ( \Rep{3}, \Rep{2}, 1)_{+4}$ & $\Rep{10}_{+4}$ & $(\Rep{6},\Rep{2},\Rep{2})$
   & + & $-$ & $-$
 \\
  $( \bar{\Rep{3}}, \Rep{1}, -4)_{+4}$ & $\Rep{10}_{+4}$ & 
$(\Rep{15},\Rep{1},\Rep{1})$
   & + & $-$ & +
 \\
  $( \Rep{1}, \Rep{1}, 6)_{+4}$ & $\Rep{10}_{+4}$ & $(\Rep{1},\Rep{1},\Rep{3})$
   & + & $-$ & +
 \\
  $( \bar{\Rep{3}},\Rep{2}, -1)_{-4}$ & $\overline{\Rep{10}}_{-4}$ &
$(\Rep{6},\Rep{2},\Rep{2})$
   & + & $-$ & $-$
 \\
  $( \Rep{3}, \Rep{1}, 4)_{-4}$ & $\overline{\Rep{10}}_{-4}$  & 
$(\Rep{15},\Rep{1},\Rep{1})$
   & + & $-$ & +
 \\
  $( \Rep{1}, \Rep{1}, -6)_{-4}$ & $\overline{\Rep{10}}_{-4}$ & 
$(\bar{\Rep{1}},\Rep{1},\Rep{3})$
   & + & $-$ & +
 \\
  \itshape \bfseries
  $( \Rep{1}, \Rep{1}, 0)_0$ & $\Rep{1}_0$ & $(\Rep{15}, \Rep{1}, \Rep{1})$
   & + & + & +
 \\
 \hline
\end{tabular}
\end{center}
\caption{Parity assignments for the components
$V_M^A = \frac{1}{2} \tr (T^A V_M) $ of the $\Rep{45}$-plet of SO(10).
$G_{SM'} = SU(3)\times SU(2)\times U(1)_Y \times U(1)_X$,
$G_{GG} = SU(5) \times U(1)$, and $G_{PS} = SU(4) \times SU(2) \times
SU(2)$.}
\label{tab:GaugeField}
\end{table}
\begin{table}[htb]
\begin{center}
\renewcommand{\arraystretch}{1.5}
\begin{tabular}{|c|c|c||cc||cc|}\hline
&&&\multicolumn{2}{ c||}{$H_d$} &
 \multicolumn{2}{ c|}{$H_u$}\\
\cline{4-7}
 $G_{SM'}$ & $G_{GG}$ & $G_{PS}$ & $Z_2^{PS}$ & $Z_2^{GG}$& $Z_2^{PS}$ &
$Z_2^{GG}$\\\hline
 $(\Rep{1},\Rep{2},-1/2)_{-2}$  &$\overline{\Rep{5}}_{-2}$ &  $(\Rep{1},\Rep{2},
\Rep{2})$ & + & +  & + & -\\
$(\Rep{1},\Rep{2},+1/2)_{+2}$  &$\Rep{5}_{+2}$ &  $(\Rep{1},\Rep{2}, \Rep{2})$ &
+ & - & + & +\\
$(\overline{\Rep{3}},\Rep{1},+1/3)_{-2}$  &$\overline{\Rep{5}}_{-2}$ & 
$(\Rep{6},\Rep{1}, \Rep{1})$ & - & + & - & -\\
$(\Rep{3},\Rep{1},-1/3)_{+2}$  &$\Rep{5}_{+2}$ &  $(\Rep{6},\Rep{1}, \Rep{1})$ &
- & - & - & +\\
\hline
\end{tabular}
\caption{Parity assignments for the components of the $\Rep{10}$
hypermultiplet.}
\label{tab:bulk10}
 \end{center}
\end{table}
\begin{table}[htb]
\begin{center}
\renewcommand{\arraystretch}{1.5}
\begin{tabular}{|c|c|c||cc||cc|}\hline
&&&\multicolumn{2}{ c||}{$\bar{\Delta}_1$} &
 \multicolumn{2}{ c|}{$\bar{\Delta}_2$}\\
\cline{4-7}
 $G_{SM'}$ & $G_{GG}$ & $G_{PS}$ & $Z_2^{PS}$ & $Z_2^{GG}$& $Z_2^{PS}$ &
$Z_2^{GG}$\\\hline
 $(\overline{\Rep{3}},\Rep{2},-1/6)_{+1}$  &$\overline{\Rep{10}}_{+1}$ & 
$(\overline{\Rep{4}},\Rep{2}, \Rep{1})$ & - & -  & + & -\\
 $(\Rep{1},\Rep{2},+1/2)_{-3}$  &$\Rep{5}_{-3}$ &  $(\overline{\Rep{4}},\Rep{2},
\Rep{1})$ & - & +  & + & +\\
 $(\Rep{3},\Rep{1},+2/3)_{+1}$  &$\overline{\Rep{10}}_{+1}$ & 
$(\Rep{4},\Rep{1}, \Rep{2})$ & + & -  & - & -\\
$\oplus(\Rep{1},\Rep{1},-1)_{+1}$&&&&&&\\
$(\Rep{3},\Rep{1},-1/3)_{-3}$  &$\Rep{5}_{-3}\oplus\Rep{1}_{+5}$ & 
$(\Rep{4},\Rep{1}, \Rep{2})$ & + & +  & - & +\\
$\oplus(\Rep{1},\Rep{1},0)_{+5}$&&&&&&\\
\hline
\end{tabular}
\caption{Parity assignments for the components of the
  $\overline{\Rep{16}}$ hypermultiplet.} 
\label{tab:bulk16}
 \end{center}
\end{table}
 The different boundary conditions break the gauge symmetry to different
subgroups at the different fixed points, concretely Pati-Salam
$\SU{4}\times\SU{2}\times\SU{2}$ at fixed point $\hat{z}_1$,
$\SU{5}\times\U{1}_X$ at $\hat{z}_3$ and flipped
$\SU{5}^\prime\times\U{1}^\prime$ at $\hat{z}_2$, while $\SO{10}$ is unbroken at
$\hat{z}_4$. The remnant gauge symmetry of the orbifold is the intersection of
the gauge symmetries at each fixed point, which is
\mbox{G$_{SM^\prime}=\SU{3}\times\SU{2}\times\U{1}_Y \times\U{1}_X$}.
The remaining U(1)$_X$ can be broken by a right-handed neutrino mass term.
All flavons, i.e.~gauge singlets, in the bulk have positive gauge parities.
Hence, the gauge and flavor breaking sectors are separated.

Analogously, by assigning the parities to the flavons living in the bulk of
the second orbifold, a zero mode is singled out and consequently the flavor
symmetry is broken. As the flavons are the bulk fields of only one
orbifold, the flavons inherit $N=2$ SUSY in 4d, which can be broken by using one
orbifold parity. If the zero mode acquires a VEV, the breaking of the symmetry
is transmitted to the fermion masses. We assume that flavons transform
non-trivially,
\begin{subequations}
\label{eq:FlavorParities}
\begin{align}
 P_1 \phi(x_\mu,z,-\hat{z}) &= \eta_1\, \phi(x_\mu,z,\hat{z}), \\
 P_2 \phi(x_\mu,z,-\hat{z}+\hat{z}_1) &= \eta_2\,
\phi(x_\mu,z,\hat{z}+\hat{z}_1), \\
 P_3 \phi(x_\mu,z,-\hat{z}+\hat{z}_3) &= \eta_3\,
\phi(x_\mu,z,\hat{z}+\hat{z}_3),
\end{align}
\end{subequations}
where the $P_{1,2,3}=Z,\,T_1 Z,\,T_2 Z$ are formed by the combination of the
translation operators $T_{1,2}$ and parity operator $Z$ acting on flavor
space. While the first parity is used to break $N=2$ SUSY to $N=1$, the
remaining two parity operators are used to generate the VEV alignment of the
flavons by singling out the appropriate zero modes.  We note that, in fact, the
flavor symmetry is already broken when the boundary conditions are introduced,
however, in order to predict the neutrino mixing, the breaking has to be
transmitted by a certain flavon VEV alignment to the fermion masses.

In general, the flavon VEV alignment is determined by a possibly complicated
flavon potential. However, in an orbifold context, (part of) the VEV alignment
can be
obtained from the action of the orbifold parities in flavor space, i.e.~the
simultaneous eigenvectors of the set of
parity operators in flavor space. Therefore, the flavon potential can remain
simple.  As the square of the parity operators has to be the
identity~\cite{Kobayashi:2008ih}, they have eigenvalues $\pm1$. We restrict
ourselves to parity operators which are within our flavor group, because other
choices of parity operators, which are not elements of the flavor group, do not
preseve the Clebsch-Gordan coefficients and would forbid e.g. the kinetic term
of the flavon field. For the
VEV alignment, we are only interested in elements which have both $+1$ and $-1$
as eigenvalues to obtain one single zero mode. VEVs originating from the same
set of parity operators are
orthogonal. However, parity operators can be chosen differently for different
representations of
$S_4$, and therefore VEVs of fields in different representations do not have to
be
orthogonal. We note that the couplings between the fields with different parity operators are forbidden because they are not invariant under the orbifold parity.

Concretely, in $S_4$, the elements of order two are in the conjugacy classes
$\mathcal{C}_{2,4}$
according to the notation in~\cite{Bazzocchi:2009pv} and its eigenvalues can
be inferred from its character. The eigenvectors to the non-degenerate
eigenvalues of each element are shown in \Tabref{tab:EVstruct}.
\begin{table}
\begin{center}
\begin{tabular}{|c|c||| c | c| c|c|}\hline
$\mathcal{C}_2$ & $3_i$ & $\mathcal{C}_4$& $3_1$ & $3_2$ & $2$\\\hline\hline
EV & $1$ &EV & $-1$ & $1$ & $1$\\\hline
 $S^2$  & $\begin{pmatrix} 1 & 1 & 1\end{pmatrix}$ &
 $T S T$  &  \multicolumn{2}{c|}{$\begin{pmatrix} -2 & 1 & 1\end{pmatrix}$} & 
$\begin{pmatrix} 1 & 1\end{pmatrix}$\\
 $T S^2 T^2$  & $\begin{pmatrix} 1 & \omega^2 & \omega \end{pmatrix}$ &
 $T^2 S$  &\multicolumn{2}{c|}{$\begin{pmatrix} -2 & \omega^2 &
\omega\end{pmatrix}$} & $\begin{pmatrix} 1 & \omega\end{pmatrix}$\\
 $S^2 T S^2 T^2$   &  $\begin{pmatrix} 1 & \omega & \omega^2\end{pmatrix}$ &
 $ S T^2$  &\multicolumn{2}{c|}{$\begin{pmatrix} -2 & \omega &
\omega^2\end{pmatrix}$} & $\begin{pmatrix} 1 & \omega^2\end{pmatrix}$\\
&& $T S T S^2$  &\multicolumn{2}{c|}{$\begin{pmatrix} 0 & -1 &
     1\end{pmatrix}$} &  $\begin{pmatrix} 1 & 1\end{pmatrix}$\\
&& $ S T S^2$  &\multicolumn{2}{c|}{$\begin{pmatrix} 0 & -\omega^2 &
     \omega\end{pmatrix}$} &  $\begin{pmatrix} 1 & \omega\end{pmatrix}$\\
&& $ S^2 T S$  &\multicolumn{2}{c|}{$\begin{pmatrix} 0 & -\omega &
     \omega^2\end{pmatrix}$} &  $\begin{pmatrix} 1 &
\omega^2\end{pmatrix}$\\\hline
\end{tabular}
\caption{Eigenvector structure of the elements of the conjugacy classes
  $\mathcal{C}_{2,4}$ in representation $2$ and $3_i$, where $S,T$ are
generators of
  $S_4$ and
  $\omega=\mathrm{e}^{2\pi\I/3}$. EV denotes a/the non-degenerate
eigenvalue and the vector in the corresponding row and column is the
eigenvector of the group element given to the eigenvalue shown at the top of the column.}
\label{tab:EVstruct}
\end{center}
\end{table}
With two parity operators, it is possible to obtain every eigenvector shown in
\Tabref{tab:EVstruct} as zero mode as well as the orthogonal complement of any
two chosen ones.
We give one example, how to obtain the VEV alignment $(1,\,1,\,1)$, which we use
in the model in the following section.
In order to obtain the VEV alignment for a triplet $\phi\sim 3_1$, we choose
\begin{align}
P_1 =& \mathbb{1}, & P_2 =& TST, &P_3 =& TSTS^2\; .
\end{align}
\begin{align*}
\mathrm{i.e.}\quad P_2=& \frac{1}{3}
\begin{pmatrix}
-1 & 2 & 2 \\
2 & 2 & -1 \\
2 & -1 & 2
\end{pmatrix} \; ,&
P_3 =&\begin{pmatrix}
1 & 0 & 0 \\
0 & 0 & 1 \\
0 & 1 & 0
\end{pmatrix} \;.
\end{align*}
As $P_1$ is the unit matrix, it does not affect the zero
  mode. Therefore, the zero mode is entirely determined by $P_{2,3}$.
$P_2$ and $P_3$ are simultaneously diagonalized by the unitary matrix $U$
\begin{equation}
U = \begin{pmatrix}
       \frac{1}{\sqrt{3}} & \frac{2}{\sqrt{6}} & 0 \\
       \frac{1}{\sqrt{3}} & -\frac{1}{\sqrt{6}} & \frac{1}{\sqrt{2}} \\
       \frac{1}{\sqrt{3}} & -\frac{1}{\sqrt{6}} & -\frac{1}{\sqrt{2}}
      \end{pmatrix} \;, \quad\quad
\begin{array}{rl}
U^{\dagger} P_2 U &= \mathrm{diag}(1,\, -1,\, 1) \;,\\
U^{\dagger} P_3 U &=  \mathrm{diag}(1,\, 1,\, -1) \;,
\end{array}
\end{equation}
The eigenvalues and eigenvectors can also be read of from
  \Tabref{tab:EVstruct}. The eigenvector to the non-degenerate
  eigenvalue $-1$ of $P_2=TST$ is $(-2,\,1,\, 1)$
    corresponding to the second column of $U$. The two
    eigenvectors to the eigenvalue $1$ are not explicitly given, but
    they are orthogonal to $(-2,\,1,\,1)$, which is sufficient for the
      analysis. Similarly, the eigenvector to the non-degenerate
      eigenvalue $-1$ of $P_3=TSTS^2$ is $(0,\, -1,\,1)$ according to \Tabref{tab:EVstruct}. This
      eigenvector determines the third column of $U$. The
      two eigenvectors to eigenvalue $+1$ are also in the orthogonal complement. Taking
      the parity assignment of $\phi$, $\eta_2=\eta_3=+1$, the eigenvectors
      to the eigenvalue $+1$ of $P_{2,3}$ are chosen and the zero mode
      lies in the intersection of the two-dimensional subspaces
      spant by the eigenvectors to the eigenvalue $+1$ of
      $P_{2,3}$. Hence, we can determine the zero mode by either
      taking the cross product of $(-2,\,1,\, 1)$ with $(0,\, -1,\,1)$ or equivalently the first column of $U$ which
        corresponds to $P_{2,3}$ having eigenvalue +1. Hence, the VEV alignment is
\begin{equation}
 \ev{\phi}= \begin{pmatrix}
                             1 & 1 & 1
                            \end{pmatrix}^T/\sqrt{3}
                            \ev{\tilde{\phi}}\; ,
\end{equation}
with $\tilde{\phi}$ being the single zero mode of $\phi$.
In order to demonstrate this, we present a model which is based on the previous
discussion.

\section{Model\label{sec:Model}}
As discussed in the previous sections, the gauge field is a bulk
  field of the two orbifolds in our setup, while the other fields are
  brane fields of the first orbifold $T^2/\mathbb{Z}_2$, leading to
  the flavor symmetry $S_4$, and can be either a brane or bulk field of the second orbifold $T^2/(\mathbb{Z}_2^I \times
\mathbb{Z}_2^{PS}\times\mathbb{Z}_2^{GG})$, which is used to break the GUT and the flavor symmetry.
Besides the gauge field, we introduce a $\Rep{16}$ brane field $\psi$, which is
localized on the $\SO{10}$ brane and leads to the SM matter, together with the $\SO{10}$ singlet fields $S_{N,\nu}$ on the SO(10) brane, the
bulk Higgs fields $H_{u,d}\sim\Rep{10}$ generating
fermion masses and breaking electroweak symmetry  as well as $\Phi\sim\Rep{45}$
and the bulk Higgs fields
 $\bar{\Delta}_i\sim\Rep{\bar{16}}$ leading to the mass terms of neutrinos.
Furthermore, there are the flavons $\phi_{1,1,1}$, $\xi$ and $\chi_{1,0,0}$ in
the bulk as well as  $\chi_{0,1,0}$ and $\chi_{0,0,1}$ on the SO(10) brane. The
index of each flavon field denotes its VEV alignment, i.e.~ $\chi_{a,b,c}\sim(a,\,b,\,c)\tilde{\chi}_{a,b,c}$.
The $\U{1}_R$ helps to implement the driving field mechanism for the flavon
potential and the additional $\mathbb{Z}_N$ charge is introduced to separate the
particles in the different sectors. 
We note that the Yukawa coupling related to a bulk zero mode have to
  be rescaled by the rescaling factor $1/\sqrt{\Lambda^2
      V}<1$, where $\Lambda$ is the cut-off of the bulk theory and
  $1/\sqrt{V}$ is the volume normalization factor of the zero
  mode as can be seen in \Appref{ModeExpand}. In our case, $V=\frac{1}{2} L_7 L_8 \sin \theta$ with $\theta$ being the angle between the
  basis vectors spanning the second orbifold and $L_{7,8}$ being their respective
  length. In principle, they are free parameters, but for concreteness, we set
  $\theta=\pi/3$ and $L_7=L_8=2 \pi R$. In the following, we will only write the zero mode of each bulk field, since we are
  interested in the low-energy spectrum.
The particle content and all charges are given in \Tabref{tab:particle
  content Model}. The orbifold parities of each Higgs field
  determines its respective zero mode and also which component obtains a VEV. In
  particular the VEVs of the $\Rep{10}$-plets are chosen to be at the
  electroweak scale. The singlet component of $\bar{\Delta}_1$ acquires a VEV
of the order of the GUT scale, whereas the VEV of the doublet component of $\bar{\Delta}_2$ is
assumed to be of the order of the electroweak scale. This additional Higgs doublet
  component and its fermionic superpartner modify the running of the
  electroweak gauge couplings, which might destroy gauge coupling
  unification. There are several possible solutions. For instance,
  corrections due to Kaluza-Klein threshold effects~\cite{Dienes:1998vh,Dienes:1998vg} might restore the unification of gauge couplings. Otherwise, it is possible to introduce additional fields either to obtain complete GUT
  representations or magic combinations, which do not spoil gauge
  coupling unification~\cite{Calibbi:2009cp}. This, however, leads to
  additional effects at the low energy scale, e.g. through new colored
  states. As we are concentrating on the flavor structure, we are
  assuming in the following that a successful gauge coupling unification can be
  obtained. The parities of $\Phi\sim\Rep{45}$ are chosen such that
  the two total SM singlets remain as zero modes, as it can be seen in \Tabref{tab:GaugeField}. In
the following, we assume, that only the component in $B-L$ direction obtains a
VEV, i.e.~$\ev{\Phi}\sim B-L$ which leads to the Georgi-Jarlskog
factor~\cite{Georgi:1979df} at the GUT scale.
\begin{table}[tbh]
\begin{center}
\begin{tabular}{|c||c||c|c|c||c|c|c|}\hline
Field & $SO(10)$ & $S_4$ & $\mathbb{Z}_N$ & $\U{1}_R$ & $\eta_I\eta_1$ &
$\eta_{PS}\eta_2$ & $\eta_{GG}\eta_3$\\\hline
$\psi$ &  $\Rep{16}$ &  $3_2$ & $0$  & $1$ & & &\\\hline
$H_u$ &  $\Rep{10}$ & $1_1$  & $-2$& $0$ & $+$ & $+$ & $-$ \\
$H_d$ &  $\Rep{10}$ & $1_1$ & $-2$ & $0$ & $+$ & $+$ & $+$ \\
$\Phi$ & $\Rep{45}$ & $1_1$  &  $4$& $0$ & $+$ & $+$ & $+$ \\
$\chi_{1,0,0}$ &  $\Rep{1}$ &  $3_1$  &  $-3$& $0$ &$+$ & $-$ & $-$ \\
$\chi_{0,0,1}$ &  $\Rep{1}$ &  $3_2$  &  $-1$& $0$ &  &  &  \\
$\chi_{0,1,0}$ &  $\Rep{1}$ &  $3_2$ & $1$ & $0$ &  &  &  \\\hline
$\bar\Delta_1$ & $\Rep{\bar{16}}$ & $1_1$ & $-1$ & $0$ & $+$ & $-$ & $-$ \\
$\bar\Delta_2$ & $\Rep{\bar{16}}$ & $1_1$ &  $3$ & $0$ & $+$ & $+$ & $-$ \\
 $S_N$ & $\Rep{1}$ & $3_2$ & $1$& $1$ & &&\\
$S_\nu$ & $\Rep{1}$ & $3_2$ & $-3$& $1$ & &&\\
$\phi_{1,1,1}$ & $\Rep{1}$ & $3_1$ & $6$ & $0$ & $+$ & $+$ & $+$ \\
$\xi$ & $\Rep{1}$ &  $1_1$ & $6$ & $0$ & $+$ & $+$ & $+$ \\\hline
\end{tabular}
\end{center}
\caption{\label{tab:particle content Model} Particle content of the
  model. Bulk fields are classified by three orbifold parities in addition to
the symmetry groups. The lower index of the flavon fields denote their zero
modes. }
\end{table}

Small neutrino masses can be generated by the seesaw mechanism starting from
\begin{equation}
 {\bf W}_\nu = \frac{y_{s}}{\sqrt{\Lambda^2 V}}  \psi \bar{\Delta}_2 S_\nu + \frac{1}{\sqrt{\Lambda^2 V}} \left(y^\nu_\phi\phi_{1,1,1}
+y^{\nu}_\xi \xi\right) S_{\nu} S_{\nu}  \;.
\end{equation}
After $\phi_{1,1,1}$ and $\xi$ obtain VEVs, $S_\nu$ becomes massive
\begin{equation}
 M_{SS}=\frac{1}{\sqrt{\Lambda^2 V}} \left( y^\nu_\phi\ev{\phi_{1,1,1}} +y^{\nu}_\xi \ev{\xi} \right)
\end{equation}
 and leads to neutrino masses
\begin{equation}
 M_{\nu} = - \frac{1}{\Lambda^2 V} (y_s \ev{\bar\Delta_2}) (M_{SS}^{-1}) (y_s \ev{\bar\Delta_2})^T
 = - m_0 U^*_\mathrm{tbm}
\begin{pmatrix}
       \frac{1}{3a+b} & 0 & 0 \\
       . & \frac{1}{b} & 0 \\
       . & . & \frac{1}{3a-b}
      \end{pmatrix} U^\dagger_\mathrm{tbm}\;,
\end{equation}
with $m_0=y_s^2\ev{\bar\Delta_2}^2/\ev{\xi}/\sqrt{\Lambda^2 V}$ as well as
$a=y^\nu_\phi\ev{\phi}/\ev{\xi}$ and
$b=y^\nu_\xi$. As $M_{SS}$ is diagonalized by the tribimaximal mixing matrix and the structure of the Yukawa couplings $y_s$ does not affect the tribimaximal mixing, the neutrino mixing matrix is of the
tribimaximal form and the neutrino masses are given by
\begin{equation}
\left(m_1,\,m_2,\,m_3\right)=m_0 \left(\frac{1}{3
a + b}, \frac{1}{b} , \frac{1}{3 a - b}\right)\;.
\end{equation}

In order to suppress the usual type-I seesaw contribution to the neutrino mass
matrix, we introduce an additional $\SO{10}$ singlet $S_N$ which combines with
the SM singlet component of $\psi$
\begin{equation}\label{eq:NuMassTerms}
{\bf W}_N = \frac{y_N}{\sqrt{\Lambda^2 V}}  \psi \bar{\Delta}_1 S_N\;,
\end{equation}
to form a pseudo-Dirac particle and there is only a small mixing with the light
neutrinos of the order of $\ev{H_u}/\ev{\bar{\Delta}_1}$  with
$\ev{\bar{\Delta}_1}$ being of the order of the GUT scale. Therefore the SM
singlet component essentially decouples from the low-energy theory. Furthermore, this term breaks the remnant $\U{1}_X$.

For clarity, we present the quark sector of the model in terms of effective
higher-dimensional operators and demonstrate in
\Appref{app:RenormalizableCouplings}, how the operators can be obtained by integrating out vector-like brane fields. The third
generation Yukawa couplings are described by
\begin{equation}\label{eq:ThirdGen}
 {\bf W}_\mathrm{3} \supset \frac{1}{\sqrt{\Lambda^2 V}}\frac{y_3^{u,d}}{M_3^2}(\psi \chi_{0,1,0})_{1_1} (\psi
\chi_{0,1,0})_{1_1} H_{u,d}\;,
\end{equation}
where $(\psi \chi_{0,1,0})_{1_1}$ denotes that the fields within the brackets
are contracted to the trivial singlet $1_1$ and $M_3$ is the mass of the
decoupled vector-like field. After the flavon $\chi_{0,1,0}$ obtains a VEV, a
mass term
\begin{equation}
 {\bf W}_\mathrm{3}\supset \frac{y_3^{u,d} \ev{\chi_{0,1,0}}^2 H_{u,d}}{\sqrt{\Lambda^2 V} M_3^2 } \psi_3^2
\end{equation}
of the third generation is generated~\cite{Dutta:2009ij,Dutta:2009bj}. The
masses of the top and bottom quark are
\begin{equation}
 m_{t,b}= \frac{ y_{3}^{u,d} \ev{\chi_{0,1,0}}^2}{\sqrt{\Lambda^2 V} M_3^2} \ev{H_{u,d}}  \;,
\end{equation}
 which requires a large top Yukawa coupling, $y_{3}^{u}$, to
  overcome the suppression from the extra-dimensional volume and the
  mass scale $M_3$. The VEV $\ev{H_{u,d}}$ is of the order of the
  electroweak scale and $y_{3}^{u,d}
  \ev{\chi_{0,1,0}}^2/(\sqrt{\Lambda^2 V} M_3^2)$ is of order one. The
  factor $m_b/m_t$ can be written in term of $\tan \beta =\ev{H_u}/\ev{H_d}$ as $m_b/m_t=y_3^d/(y_3^u \tan \beta)$.

Similarly, the Yukawa couplings for the first and second generation can be
obtained from
\begin{equation}\label{eq:FirstTwoGen}
 {\bf W}_\mathrm{12} \supset \frac{1}{(\sqrt{\Lambda^2 V})^2} \left( \frac{y_2^{u,d}}{M_2^3}(\psi \chi_{0,0,1})_{1_1} (\psi
\chi_{0,0,1})_{1_1} \Phi H_{u,d}
 +\frac{y_{12}^{u,d}}{\sqrt{\Lambda^2 V} M_{12}^3} (\psi\chi_{1,0,0})_{1_1}(\psi\chi_{0,0,1})_{1_1} \xi
H_{u,d} \right) \; .
\end{equation}
The scale, at which the operators are generated, is denoted by $M_2$
and $M_{12}$. 
As discussed at the beginning of the section, $\Phi\sim\Rep{45}$
obtains a VEV in $B-L$ direction leading to the Georgi-Jarlskog
factor~\cite{Georgi:1979df} at the GUT scale.
This contributes to the mass matrices subdominantly and leads to the mass of the
first generations as well as the Cabibbo angle.
The charged fermion mass matrices are given by
\begin{align}
M_{u,d} =&\, m_{t,b} \begin{pmatrix}
       0 & a^{u,d} & 0  \\
       a^{u,d} & b^{u,d} & 0\\
       0 & 0 & 1
      \end{pmatrix}\;,  &
M_{l} =&\, m_b \begin{pmatrix}
       0 & a^d & 0  \\
       a^d & 3\, b^d & 0\\
       0 & 0 & 1
      \end{pmatrix}\label{eq:chargedLepton}\; ,
    \end{align}
where 
\begin{align}
a^{u,d}&=\frac{1}{(\sqrt{\Lambda^2 V})^2} \frac{y_{12}^{u,d}}{y_{3}^{u,d}}
\frac{M_3^2
  \ev{\chi_{1,0,0}}\ev{\chi_{0,0,1}}\ev{\xi}}{M_{12}^3\ev{\chi_{0,1,0}}^2}\quad\quad\mathrm{and}&\quad
b^{u,d}&=\frac{1}{\sqrt{\Lambda^2 V}} \frac{y_2^{u,d}}{y_3^{u,d}}
\frac{ M_3^2 \ev{\chi_{0,0,1}}^2 \ev{\Phi}}{M_2^3 \ev{\chi_{0,1,0}}^2}
\end{align}
with $a^{u,d}$ being naturally smaller than $b^{u,d}$ through the
additional suppression by the volume factor. The Cabibbo angle $\theta_c$ is approximately given by
$\sin\theta_c=a^d/b^d-a^u/b^u$. Hence, the dominant contribution to the Cabibbo angle is from the down-type quark mass matrix
\begin{equation}
0.04-0.05\simeq\sqrt{\frac{m_u}{m_c}}=\frac{a^u}{b^u}\ll \frac{a^d}{b^d}=\sqrt{\frac{m_d}{m_s}}\simeq 0.23\;.
\end{equation}
We note that the other quark mixing angles can be obtained from higher order corrections.
As the structure of the charged lepton mass matrix in \Eqref{eq:chargedLepton}
is connected to the down-type mass matrix $M_d$ and the CKM mixing is mainly
generated by $M_d$, there is a correction to the tribimaximal mixing matrix~\cite{King:2005bj,*Masina:2005hf,*Antusch:2005kw}
which results in a small deviation from the tribimaximal mixing
\begin{align}
s_{12}^2&=\frac13-\frac{2\theta_c}{9}+\frac{\theta_c^2}{54},&
s_{13}^2&=\frac{\theta_c^2}{18},&
s_{23}^2&=\frac12-\frac{\theta_c^2}{36}
\end{align}
with $s_{ij}=\sin\theta_{ij}$. Note that the solar mixing angle $\theta_{12}$ is corrected towards smaller values.
Depending on the absolute mass scale of neutrino masses, the angles are further
corrected by the renormalization group evolution, which we can neglect in case of a hierarchical spectrum~\cite{Antusch:2003kp}.

Finally, we discuss the flavon VEV alignment and show that we obtain the
required one. The VEV alignment of the bulk flavon fields is readily obtained
from the boundary conditions, which project out one single zero mode. However,
we still have to show that the zero mode develops a VEV and the VEV alignment of
the brane flavon fields is achieved. The alignment of $\phi_{1,1,1}$ has been
demonstrated in the previous section.
In order to obtain the VEV alignment for the triplet $\chi_{1,0,0}\sim3_2$, we
choose the parity
operators,  in flavor space as
\begin{align}
P_1 =& \mathbb{1}, & P_2 =& STS^2 & P_3 =& TSTS^2\; .
\end{align}
which leads to
\begin{equation}
\ev{\chi_{1,0,0}}=\begin{pmatrix}
                             1 & 0 & 0
                            \end{pmatrix}^T\ev{\tilde{\chi}_{1,0,0}}
\end{equation}
with $\tilde\chi_{1,0,0}$ denoting the single zero mode of flavon $\chi_{1,0,0}$
and its parities are given in \Tabref{tab:particle content Model}.

The remaining VEV alignment can be obtained
from the flavon potential by using the driving field mechanism
\begin{multline}
 {\bf W}_{fl}=\sum_i\delta_0^{(i)} \left(\lambda_{ij}  X_j
-M_i^2\right)\\
+  \frac{\delta_{-12}}{(\sqrt{\Lambda^2 V})^2}  (\alpha_1 \xi^2-\alpha_2\phi_{1,1,1}^2)
+ \frac{\alpha_3 \delta_{4}}{\sqrt{\Lambda^2
    V}}\chi_{1,0,0}\chi_{0,0,1}+ \alpha_4 \delta_{-2}\chi_{0,1,0}^2
+  \delta_{2}^{(i)}\kappa_{ij} Y_j
\end{multline}
with 
\begin{equation}
X=\left(\frac{\bar{\xi}\xi}{(\sqrt{\Lambda^2
      V})^2}\,,\,\frac{\chi_{1,0,0}^2}{(\sqrt{\Lambda^2
      V})^2}\,,\,\chi_{0,0,1}\chi_{0,1,0}\right)^T\quad\quad
\mathrm{and}\quad\quad 
Y=\left(\frac{1}{\sqrt{\Lambda^2
      V}}\chi_{1,0,0}\chi_{0,1,0}\,,\,\chi_{0,0,1}^2\right)^T\;.
\end{equation}
The
matrices $\left(\lambda_{ij}\right)$ and $\left(\kappa_{ij}\right)$ are
non-singular as well as the couplings $\alpha_i\neq0$ and $M_i\neq0$. We introduced one additional flavon field
$\bar\xi$ to form a singlet $\bar\xi\xi$ and several driving fields
$\delta_n^{(i)}$. The index of each driving field denotes its $\mathbb{Z}_N$
charge. They are listed in \Tabref{tab:DrivingFields}.
\begin{table}[tbh]
\begin{center}
\begin{tabular}{|c||c||c|c|c||c|c|c|}\hline
Field & $SO(10)$ & $S_4$ & $\mathbb{Z}_N$ & $\U{1}_R$ &$\eta_I\eta_1$ &
$\eta_{PS}\eta_2$ & $\eta_{GG}\eta_3$\\\hline
$\bar{\xi}$ &  $\Rep{1}$ &  $1$ & $-6$ & $0$ & $+$ & $+$ & $+$ \\\hline
$\delta_0^{(i)}$ & $\Rep{1}$ &  $1_1$ & $0$ & $2$ & & &  \\
$\delta_2^{(i)}$ & $\Rep{1}$ &  $1_1$ & $2$ & $2$ & & &  \\
$\delta_{-2}$ & $\Rep{1}$ &  $1_1$ & $-2$ & $2$ & & &  \\
$\delta_{4}$ & $\Rep{1}$ &  $1_1$ & $4$ & $2$ & & &  \\
$\delta_{-12}$ & $\Rep{1}$ &  $1_1$ & $-12$ & $2$ & & &  \\\hline
\end{tabular}
\end{center}
\caption{\label{tab:DrivingFields} The additional flavon field $\bar{\xi}$ as
well as driving fields $\delta_n^{(i)}$ on the $\SO{10}$ brane which are needed
to obtain the relevant VEV alignment. The lower index of $\delta$ denotes its
$\mathbb{Z}_N$ charge and a possible upper index labels different driving fields
with the same $\mathbb{Z}_N$ charge.}
\end{table}
In the SUSY limit, the minima of the flavon potential can be obtained from the
condition of vanishing $F$-terms
\begin{equation}
F_\delta=\frac{\partial {\bf W}_\mathrm{fl}}{\partial\delta} \stackrel{!}{=} 0
\end{equation}
with $\delta$ being one of the driving (or flavon) fields. The $F$ term conditions
of the flavon fields are readily satisfied by the vanishing of all driving field
VEVs. The $F$ term conditions of the driving fields determine the flavon VEVs.
 All components of $X$ obtain a VEV due to the non-singularity of
$\left(\lambda\right)_{ij}$. The $F$ term condition $F_{\delta_4}=0$ forces the
VEV in the first component of $\ev{\chi_{0,0,1}}$ to vanish. Similarly, the
non-singularity of $\left(\kappa_{ij}\right)$ leads to a vanishing of the first
component of the VEV of $\ev{\chi_{0,1,0}}$ as well as the product
$\ev{\chi_{0,0,1}}_2\ev{\chi_{0,0,1}}_3$. Therefore, either the third or second
component of $\ev{\chi_{0,0,1}}$ vanishes. We choose the second component.
Furthermore, $F_{\delta_{-2}}=0$ together with the non-vanishing of
$\ev{\chi_{0,0,1}\chi_{0,1,0}}$ leads to the vanishing of the third component of
$\ev{\chi_{0,1,0}}$.  The new fields do not introduce additional couplings up to
a certain order depending on the $\mathbb{Z}_N$ charge.

\section{Conclusions}
In this paper, we investigated a model, where an $S_4$ flavor symmetry arises
from an orbifold compactification and it is broken by the boundary conditions
together with the $\SO{10}$ gauge symmetry on another orbifold. All possible VEV
alignments by only using elements of $S_4$ as parity
operators have been summarized in
\Tabref{tab:EVstruct}. Finally, we gave an example in the context of
$\SO{10}\times S_4$ which leads to a phenomenologically viable neutrino mass
matrix as well as enables to fit the masses of quarks and charged leptons. The model predicts the tribimaximal mixing at leading order 
in the lepton sector and the Gatto-Sartori-Tonin relation~\cite{Gatto:1968ss,*Oakes:1969vm} in the quark sector, which leads to a quantitatively correct Cabibbo mixing angle. The
solar mixing angle $\theta_{12}$ receives a small correction from the charged leptons towards a smaller value and a small $\theta_{13}$ is
generated. The VEV alignment obtained from the orbifold is an
essential ingredient to obtain the required VEV alignment.
We state that the VEV alignment mechanism can also be used for flavor symmetries
arising from different orbifolds~\cite{Adulpravitchai:2009id} as well as for
models with flavored Higgs fields. \\

\begin{acknowledgments}
We would like to thank J.~Jaeckel, M.~Lindner, M.~Ratz and K.~Schmidt-Hoberg for
useful discussions, as well as
F. Br\"ummer for comments on the first version. AA would like to thank
F.~Feruglio and H.~Zhang for useful discussions. AA has been supported by the
DFG-Sonderforschungsbereich Transregio 27.
\end{acknowledgments}

\appendix

\section{Generation of Effective Operators\label{app:RenormalizableCouplings}}
In this section, we demonstrate how to obtain the effective operators, which
have been used to generate the charged fermion mass matrices in
\Secref{sec:Model}. The vector-like brane fields, which are integrated out, are
summarized in \Tabref{tab:vector-like}.
\begin{table}[tbh]
\begin{center}
\begin{tabular}{|c||c||c|c|}\hline
Field & $SO(10)_{\mathbb{Z}_N}$ & $S_4$ &  $\U{1}_R$\\\hline
$\Psi_{3}\oplus\bar\Psi_3$ &  $\Rep{16}_1\oplus\Rep{\bar{16}}_{-1}$ & $1_1$  &
$1$ \\
 $\Psi_{2} \oplus\bar\Psi_2$ & $\Rep{16}_{-1}\oplus\Rep{\bar{16}}_1$ & $1_1$ &  
$1$\\
$\Psi_{2}'\oplus\bar\Psi_2'$ & $\Rep{16}_3\oplus\Rep{\bar{16}}_{-3}$ & $1_1$ & 
$1$ \\
$\Psi_{12}\oplus\bar\Psi_{12}$ & $\Rep{16}_{-3}\oplus\Rep{\bar{16}}_{3}$ & $1_1$  & $1$\\
$\Psi_{12}'\oplus\bar\Psi_{12}'$ & $\Rep{16}_5\oplus\Rep{\bar{16}}_{-5}$ & $1_1$  & $1$
\\\hline
\end{tabular}
\end{center}
\caption{\label{tab:vector-like} Vector-like brane fields which generate the
required effective operators. }
\end{table}
The effective term which leads to the third generation masses in
\Eqref{eq:ThirdGen} can be obtained from
\begin{equation}
 {\bf W}_{\mathrm{LO}3} =h_3(\psi \chi_{0,1,0}) \bar{\Psi}_{3} + \frac{h_{u,d}}{\sqrt{\Lambda^2 V}} 
\Psi_{3} H_{u,d} \Psi_{3} + M_{\Psi 3} \bar{\Psi}_{3} \Psi_{3}\;.
\end{equation}
The scale $M_3^2$ is therefore given by $M_{\Psi3}^2$.
The first term in \Eqref{eq:FirstTwoGen}, which generates the mass of the second
generation is obtained by integrating out two vector-like fields
\begin{equation}\label{eq:SecondGenRen}
 {\bf W}_{\mathrm{LO}2} = h_2  (\psi \chi_{0,0,1}) \bar{\Psi}_{2}   + \frac{h_{45}}{\sqrt{\Lambda^2 V}}  
\Psi_2 \Phi \bar{\Psi}_{2}' +  \frac{h_{u,d}'}{\sqrt{\Lambda^2 V}} \Psi_2  H_{u,d} \Psi_2'
 + M_{\Psi2} \bar{\Psi}_{2} \Psi_{2} + M_{\Psi 2'} \bar{\Psi}_{2}' \Psi_2' \;.
\end{equation}
After integrating out, $\Psi_2'+\bar\Psi_2'$, we obtain the effective dimension
4 operator
\begin{equation}
{\bf W}_{d=4}\supset \frac{h_{u,d}^\prime h_{45}}{(\sqrt{\Lambda^2 V})^2 M_{\Psi2'} } \Psi_{2} \Phi
H_{u,d} \Psi_{2}\;,
\end{equation}
which leads to \Eqref{eq:FirstTwoGen} after integrating out $\Psi_2+\bar\Psi_2$. Hence, the scale of the operator in \Eqref{eq:FirstTwoGen} 
$M_2^3$ equals $M_{\Psi2}^2M_{\Psi2'}$.
Finally, the Cabibbo mixing angle can be generated from
\begin{equation}\label{eq:CabibboRen}
{\bf W}_{\mathrm{LO}12} = \frac{h_{12}}{\sqrt{\Lambda^2 V}}  (\psi \chi_{1,0,0}) \bar{\Psi}_{12} +
\frac{h_{u,d}^{\prime\prime}}{\sqrt{\Lambda^2 V}}  \Psi_{12} H_{u,d} \Psi_{12}' + \frac{h_\xi}{\sqrt{\Lambda^2 V}} 
\bar\Psi_{12}'\Psi_2\xi
  + M_{\Psi12'} \bar{\Psi}_{12}' \Psi_{12}' 
\end{equation}
in combination with \Eqref{eq:SecondGenRen}. The second term in
\Eqref{eq:FirstTwoGen} is then generated in two steps. After integrating out
$\Psi_{12}' + \bar\Psi_{12}'$, the effective dimension 4 operator
\begin{equation}
{\bf W}_{d=4} \supset \frac{h_{u,d}^{\prime\prime} h_\xi}{(\sqrt{\Lambda^2 V})^2 M_{\Psi12}' } \Psi_{12}
H_{u,d} \Psi_{2} \xi  
\end{equation}
is generated and a subsequent decoupling of $\Psi_{12}+\bar\Psi_{12}$ and
$\Psi_2+\bar\Psi_2$ leads to the required term with $M_{12}^3=M_{\Psi2}M_{\Psi12}M_{\Psi12'}$.
\section{Mode Expansion} \label{ModeExpand}
The possible boundary conditions of functions on the
orbifold $T^2/(Z_2^I \times Z_2^{PS} \times Z_2^{GG})$ are characterized by
three parities ($a,b = +,-$),
\begin{eqnarray}
\phi_{\pm a b}(x,z,-\hat{z}) &=& \pm \phi_{\pm a b}(x,z,\hat{z})\;,
\nonumber\\
\phi_{a \pm b}(x,z,-\hat{z}+\hat{z}_1) &=& \pm \phi_{a \pm
b}(x,z,\hat{z}+\hat{z}_1)\;,
\nonumber\\
\phi_{a b \pm}(x,z,-\hat{z}+\hat{z}_3) &=& \pm \phi_{a b
\pm}(x,z,\hat{z}+\hat{z}_3)\;. \label{boundary1234}
\end{eqnarray}
The mode expansion of functions with the boundary conditions reads
\begin{subequations}
\begin{align}
\phi_{+++}(x,z,\hat{z}) &=\frac{2}{\sqrt{2^{\delta_{n,0}\delta_{m,0}} L_7 L_8
\sin \theta}}
 \left[ \delta_{0,m} \sum_{n=0}^{\infty}+
\sum_{m=1}^{\infty}\sum_{n=-\infty}^{\infty} \right]
\phi_{+++}^{(2m,2n)}(x,z) \nonumber \\&\hspace{3.96cm} \times 
\cos\left(\frac{2m \pi}{L_7}(x_7-\frac{x_8}{\tan
\theta})+\frac{2n \pi}{L_8 \sin \theta}x_8\right) ,
 \\
\phi_{++-}(x,z,\hat{z}) &=\frac{2}{\sqrt{ L_7 L_8
\sin \theta}}
 \left[ \delta_{0,m} \sum_{n=0}^{\infty}+
\sum_{m=1}^{\infty}\sum_{n=-\infty}^{\infty} \right]
\phi_{++-}^{(2m,2n+1)}(x,z)\nonumber \\&\hspace{3.96cm} \times
\cos\left(\frac{2m \pi}{L_7}(x_7-\frac{x_8}{\tan
\theta})+\frac{(2n+1) \pi}{L_8 \sin \theta}x_8\right) ,
 \\
\phi_{+-+}(x,z,\hat{z}) &=\frac{2}{\sqrt{ L_7 L_8
\sin \theta}}
 \left[
\sum_{m=0}^{\infty}\sum_{n=-\infty}^{\infty} \right]
\phi_{+-+}^{(2m+1,2n)}(x,z)\nonumber \\&\hspace{3.96cm} \times
\cos\left(\frac{(2m+1) \pi}{L_7}(x_7-\frac{x_8}{\tan
\theta})+\frac{2n \pi}{L_8 \sin \theta}x_8\right) ,
\\
\phi_{+--}(x,z,\hat{z}) &=\frac{2}{\sqrt{ L_7 L_8
\sin \theta}}
 \left[
\sum_{m=0}^{\infty}\sum_{n=-\infty}^{\infty} \right]
\phi_{+--}^{(2m+1,2n+1)}(x,z)\nonumber \\&\hspace{3.96cm} \times
\cos\left(\frac{(2m+1) \pi}{L_7}(x_7-\frac{x_8}{\tan
\theta})+\frac{(2n+1) \pi}{L_8 \sin \theta}x_8\right) ,
\\
\phi_{-++}(x,z,\hat{z}) &=\frac{2}{\sqrt{ L_7 L_8
\sin \theta}}
 \left[
\sum_{m=0}^{\infty}\sum_{n=-\infty}^{\infty} \right]
\phi_{-++}^{(2m+1,2n+1)}(x,z)
\nonumber \\&\hspace{3.96cm} \times
\sin\left(\frac{(2m+1) \pi}{L_7}(x_7-\frac{x_8}{\tan \theta})+\frac{(2n+1)
\pi}{L_8 \sin \theta}x_8\right) ,
\\
\phi_{-+-}(x,z,\hat{z}) &=\frac{2}{\sqrt{ L_7 L_8
\sin \theta}}
 \left[
\sum_{m=0}^{\infty}\sum_{n=-\infty}^{\infty} \right]
\phi_{-+-}^{(2m+1,2n)}(x,z)
\nonumber \\&\hspace{3.96cm} \times
\sin\left(\frac{(2m+1) \pi}{L_7}(x_7-\frac{x_8}{\tan \theta})+\frac{2n \pi}{L_8
\sin \theta}x_8\right) ,
\end{align}
\begin{align}
\phi_{--+}(x,z,\hat{z}) &=\frac{2}{\sqrt{ L_7 L_8
\sin \theta}}
 \left[ \delta_{0,m} \sum_{n=0}^{\infty}+
\sum_{m=1}^{\infty}\sum_{n=-\infty}^{\infty} \right]
\phi_{++-}^{(2m,2n+1)}(x,z)
\nonumber \\&\hspace{3.96cm} \times
\sin\left(\frac{2m \pi}{L_7}(x_7-\frac{x_8}{\tan \theta})+\frac{(2n+1) \pi}{L_8
\sin \theta}x_8\right) ,
 \\
\phi_{---}(x,z,\hat{z}) &=\frac{2}{\sqrt{ L_7 L_8
\sin \theta}}
 \left[ \delta_{0,m} \sum_{n=0}^{\infty}+
\sum_{m=1}^{\infty}\sum_{n=-\infty}^{\infty} \right]
\phi_{---}^{(2m,2n)}(x,z)
\nonumber \\&\hspace{3.96cm} \times
\sin\left(\frac{2m \pi}{L_7}(x_7-\frac{x_8}{\tan \theta})+\frac{2n \pi}{L_8 \sin
\theta}x_8\right),
\end{align}
\end{subequations}
with $\theta$ being the angle between the basis vectors spanning
  the orbifold $T^2/(Z_2^I \times Z_2^{PS} \times Z_2^{GG})$ and
  $L_{7,8}$ being their respective length. The wave functions are normalized
by the volume of the fundamental region.


\bibliography{Orbifold}

\providecommand{\href}[2]{#2}\begingroup\raggedright\begin{thebibliography}{10}

\bibitem{Harrison:2002er}
P.~F. Harrison, D.~H. Perkins, and W.~G. Scott, {\it {Tri-bimaximal mixing and
  the neutrino oscillation data}},  {\em Phys. Lett.} {\bf B530} (2002) 167,
  [\href{http://xxx.lanl.gov/abs/hep-ph/0202074}{{\tt hep-ph/0202074}}].

\bibitem{Adulpravitchai:2009kd}
A.~Adulpravitchai, A.~Blum, and M.~Lindner, {\it {Non-Abelian Discrete Groups
  from the Breaking of Continuous Flavor Symmetries}},  {\em JHEP} {\bf 09}
  (2009) 018, [\href{http://xxx.lanl.gov/abs/0907.2332}{{\tt
  arXiv:0907.2332}}].

\bibitem{Berger:2009tt}
J.~Berger and Y.~Grossman, {\it {Model of leptons from $SO(3) \to A_4$}},  {\em
  JHEP} {\bf 02} (2010) 071, [\href{http://xxx.lanl.gov/abs/0910.4392}{{\tt
  arXiv:0910.4392}}].

\bibitem{Kobayashi:2004ya}
T.~Kobayashi, S.~Raby, and R.-J. Zhang, {\it {Searching for realistic 4d string
  models with a Pati-Salam symmetry: Orbifold grand unified theories from
  heterotic string compactification on a Z(6) orbifold}},  {\em Nucl. Phys.}
  {\bf B704} (2005) 3--55, [\href{http://xxx.lanl.gov/abs/hep-ph/0409098}{{\tt
  hep-ph/0409098}}].

\bibitem{Kobayashi:2006wq}
T.~Kobayashi, H.~P. Nilles, F.~Ploger, S.~Raby, and M.~Ratz, {\it {Stringy
  origin of non-Abelian discrete flavor symmetries}},  {\em Nucl. Phys.} {\bf
  B768} (2007) 135--156, [\href{http://xxx.lanl.gov/abs/hep-ph/0611020}{{\tt
  hep-ph/0611020}}].

\bibitem{Ko:2007dz}
P.~Ko, T.~Kobayashi, J.-h. Park, and S.~Raby, {\it {String-derived D4 flavor
  symmetry and phenomenological implications}},  {\em Phys. Rev.} {\bf D76}
  (2007) 035005, [\href{http://xxx.lanl.gov/abs/0704.2807}{{\tt
  arXiv:0704.2807}}].

\bibitem{Abe:2009vi}
H.~Abe, K.-S. Choi, T.~Kobayashi, and H.~Ohki, {\it {Non-Abelian Discrete
  Flavor Symmetries from Magnetized/Intersecting Brane Models}},  {\em Nucl.
  Phys.} {\bf B820} (2009) 317--333,
  [\href{http://xxx.lanl.gov/abs/0904.2631}{{\tt arXiv:0904.2631}}].

\bibitem{Abe:2010ii}
H.~Abe, K.-S. Choi, T.~Kobayashi, and H.~Ohki, {\it {Flavor structure from
  magnetic fluxes and non-Abelian Wilson lines}},  {\em Phys. Rev.} {\bf D81}
  (2010) 126003, [\href{http://xxx.lanl.gov/abs/1001.1788}{{\tt
  arXiv:1001.1788}}].

\bibitem{Dixon:1986qv}
L.~J. Dixon, D.~Friedan, E.~J. Martinec, and S.~H. Shenker, {\it {The Conformal
  Field Theory of Orbifolds}},  {\em Nucl. Phys.} {\bf B282} (1987) 13--73.

\bibitem{Watari:2002fd}
T.~Watari and T.~Yanagida, {\it {Geometric origin of large lepton mixing in a
  higher dimensional spacetime}},  {\em Phys. Lett.} {\bf B544} (2002)
  167--175, [\href{http://xxx.lanl.gov/abs/hep-ph/0205090}{{\tt
  hep-ph/0205090}}].

\bibitem{Watari:2002tf}
T.~Watari and T.~Yanagida, {\it {Higher dimensional supersymmetry as an origin
  of the three families for quarks and leptons}},  {\em Phys. Lett.} {\bf B532}
  (2002) 252--258, [\href{http://xxx.lanl.gov/abs/hep-ph/0201086}{{\tt
  hep-ph/0201086}}].

\bibitem{Altarelli:2006kg}
G.~Altarelli, F.~Feruglio, and Y.~Lin, {\it {Tri-bimaximal neutrino mixing from
  orbifolding}},  {\em Nucl. Phys.} {\bf B775} (2007) 31--44,
  [\href{http://xxx.lanl.gov/abs/hep-ph/0610165}{{\tt hep-ph/0610165}}].

\bibitem{Adulpravitchai:2009id}
A.~Adulpravitchai, A.~Blum, and M.~Lindner, {\it {Non-Abelian Discrete Flavor
  Symmetries from $T^2/Z_N$ Orbifolds}},  {\em JHEP} {\bf 07} (2009) 053,
  [\href{http://xxx.lanl.gov/abs/0906.0468}{{\tt arXiv:0906.0468}}].

\bibitem{Kawamura:2000ev}
Y.~Kawamura, {\it {Triplet-doublet splitting, proton stability and extra
  dimension}},  {\em Prog. Theor. Phys.} {\bf 105} (2001) 999--1006,
  [\href{http://xxx.lanl.gov/abs/hep-ph/0012125}{{\tt hep-ph/0012125}}].

\bibitem{Burrows:2009pi}
T.~J. Burrows and S.~F. King, {\it {$A_4$ Family Symmetry from SU(5) SUSY GUTs
  in 6d}},  {\em Nucl. Phys.} {\bf B835} (2010) 174--196,
  [\href{http://xxx.lanl.gov/abs/0909.1433}{{\tt arXiv:0909.1433}}].

\bibitem{Scherk:1978ta}
J.~Scherk and J.~H. Schwarz, {\it {Spontaneous Breaking of Supersymmetry
  Through Dimensional Reduction}},  {\em Phys. Lett.} {\bf B82} (1979) 60.

\bibitem{Scherk:1979zr}
J.~Scherk and J.~H. Schwarz, {\it {How to Get Masses from Extra Dimensions}},
  {\em Nucl. Phys.} {\bf B153} (1979) 61--88.

\bibitem{Ibanez:1986tp}
L.~E. Ibanez, H.~P. Nilles, and F.~Quevedo, {\it {Orbifolds and Wilson Lines}},
   {\em Phys. Lett.} {\bf B187} (1987) 25--32.

\bibitem{Haba:2006dz}
N.~Haba, A.~Watanabe, and K.~Yoshioka, {\it {Twisted flavors and tri/bi-maximal
  neutrino mixing}},  {\em Phys. Rev. Lett.} {\bf 97} (2006) 041601,
  [\href{http://xxx.lanl.gov/abs/hep-ph/0603116}{{\tt hep-ph/0603116}}].

\bibitem{Seidl:2008yf}
G.~Seidl, {\it {Unified model of fermion masses with Wilson line flavor
  symmetry breaking}},  {\em Phys. Rev.} {\bf D81} (2010) 025004,
  [\href{http://xxx.lanl.gov/abs/0811.3775}{{\tt arXiv:0811.3775}}].

\bibitem{Kobayashi:2008ih}
T.~Kobayashi, Y.~Omura, and K.~Yoshioka, {\it {Flavor Symmetry Breaking and
  Vacuum Alignment on Orbifolds}},  {\em Phys. Rev.} {\bf D78} (2008) 115006,
  [\href{http://xxx.lanl.gov/abs/0809.3064}{{\tt arXiv:0809.3064}}].

\bibitem{Pakvasa:1978tx}
S.~Pakvasa and H.~Sugawara, {\it {Mass of the t Quark in SU(2) x U(1)}},  {\em
  Phys. Lett.} {\bf B82} (1979) 105.

\bibitem{Yamanaka:1981pa}
Y.~Yamanaka, H.~Sugawara, and S.~Pakvasa, {\it Permutation symmetries and the
  fermion mass matrix},  {\em Phys. Rev.} {\bf D25} (1982) 1895.

\bibitem{Brown:1984mq}
T.~Brown, N.~Deshpande, S.~Pakvasa, and H.~Sugawara, {\it Cp nonconservation
  and rare processes in s(4) model of permutation symmetry},  {\em Phys. Lett.}
  {\bf B141} (1984) 95.

\bibitem{Brown:1984dk}
T.~Brown, S.~Pakvasa, H.~Sugawara, and Y.~Yamanaka, {\it Neutrino masses,
  mixing and oscillations in s(4) model of permutation symmetry},  {\em Phys.
  Rev.} {\bf D30} (1984) 255.

\bibitem{Lee:1994qx}
D.-G. Lee and R.~N. Mohapatra, {\it {An SO(10) x S(4) scenario for naturally
  degenerate neutrinos}},  {\em Phys. Lett.} {\bf B329} (1994) 463--468,
  [\href{http://xxx.lanl.gov/abs/hep-ph/9403201}{{\tt hep-ph/9403201}}].

\bibitem{Ma:2005pd}
E.~Ma, {\it {Neutrino mass matrix from S(4) symmetry}},  {\em Phys. Lett.} {\bf
  B632} (2006) 352--356, [\href{http://xxx.lanl.gov/abs/hep-ph/0508231}{{\tt
  hep-ph/0508231}}].

\bibitem{Hagedorn:2006ug}
C.~Hagedorn, M.~Lindner, and R.~N. Mohapatra, {\it {S(4) flavor symmetry and
  fermion masses: Towards a grand unified theory of flavor}},  {\em JHEP} {\bf
  06} (2006) 042, [\href{http://xxx.lanl.gov/abs/hep-ph/0602244}{{\tt
  hep-ph/0602244}}].

\bibitem{Cai:2006mf}
Y.~Cai and H.-B. Yu, {\it {An SO(10) GUT Model with $S4$ Flavor Symmetry}},
  {\em Phys. Rev.} {\bf D74} (2006) 115005,
  [\href{http://xxx.lanl.gov/abs/hep-ph/0608022}{{\tt hep-ph/0608022}}].

\bibitem{Caravaglios:2006aq}
F.~Caravaglios and S.~Morisi, {\it Gauge boson families in grand unified
  theories of fermion masses: {$E_6^4\times S_4$}},  {\em Int. J. Mod. Phys.}
  {\bf A22} (2007) 2469--2492,
  [\href{http://xxx.lanl.gov/abs/hep-ph/0611078}{{\tt hep-ph/0611078}}].

\bibitem{Zhang:2006fv}
H.~Zhang, {\it {Flavor S(4) x Z(2) symmetry and neutrino mixing}},  {\em Phys.
  Lett.} {\bf B655} (2007) 132--140,
  [\href{http://xxx.lanl.gov/abs/hep-ph/0612214}{{\tt hep-ph/0612214}}].

\bibitem{Koide:2007sr}
Y.~Koide, {\it {$S_4$ Flavor Symmetry Embedded into SU(3) and Lepton Masses and
  Mixing}},  {\em JHEP} {\bf 08} (2007) 086,
  [\href{http://xxx.lanl.gov/abs/0705.2275}{{\tt arXiv:0705.2275}}].

\bibitem{Parida:2008pu}
M.~K. Parida, {\it {Intermediate left-right gauge symmetry, unification of
  couplings and fermion masses in SUSY $SO(10)\times S_4$}},  {\em Phys. Rev.}
  {\bf D78} (2008) 053004, [\href{http://xxx.lanl.gov/abs/0804.4571}{{\tt
  arXiv:0804.4571}}].

\bibitem{Lam:2008sh}
C.~S. Lam, {\it {The Unique Horizontal Symmetry of Leptons}},  {\em Phys. Rev.}
  {\bf D78} (2008) 073015, [\href{http://xxx.lanl.gov/abs/0809.1185}{{\tt
  arXiv:0809.1185}}].

\bibitem{Bazzocchi:2008ej}
F.~Bazzocchi and S.~Morisi, {\it {S4 as a natural flavor symmetry for lepton
  mixing}},  {\em Phys. Rev.} {\bf D80} (2009) 096005,
  [\href{http://xxx.lanl.gov/abs/0811.0345}{{\tt arXiv:0811.0345}}].

\bibitem{Ishimori:2008fi}
H.~Ishimori, Y.~Shimizu, and M.~Tanimoto, {\it {S4 Flavor Symmetry of Quarks
  and Leptons in SU(5) GUT}},  {\em Prog. Theor. Phys.} {\bf 121} (2009)
  769--787, [\href{http://xxx.lanl.gov/abs/0812.5031}{{\tt arXiv:0812.5031}}].

\bibitem{Bazzocchi:2009da}
F.~Bazzocchi, L.~Merlo, and S.~Morisi, {\it {Phenomenological Consequences of
  See-Saw in S4 Based Models}},  {\em Phys. Rev.} {\bf D80} (2009) 053003,
  [\href{http://xxx.lanl.gov/abs/0902.2849}{{\tt arXiv:0902.2849}}].

\bibitem{Altarelli:2009gn}
G.~Altarelli, F.~Feruglio, and L.~Merlo, {\it {Revisiting Bimaximal Neutrino
  Mixing in a Model with S4 Discrete Symmetry}},  {\em JHEP} {\bf 05} (2009)
  020, [\href{http://xxx.lanl.gov/abs/0903.1940}{{\tt arXiv:0903.1940}}].

\bibitem{Ishimori:2009ns}
H.~Ishimori, Y.~Shimizu, and M.~Tanimoto, {\it {S4 Flavor Model of Quarks and
  Leptons}},  {\em Prog. Theor. Phys. Suppl.} {\bf 180} (2010) 61--71,
  [\href{http://xxx.lanl.gov/abs/0904.2450}{{\tt arXiv:0904.2450}}].

\bibitem{Grimus:2009pg}
W.~Grimus, L.~Lavoura, and P.~O. Ludl, {\it {Is S4 the horizontal symmetry of
  tri-bimaximal lepton mixing?}},  {\em J. Phys.} {\bf G36} (2009) 115007,
  [\href{http://xxx.lanl.gov/abs/0906.2689}{{\tt arXiv:0906.2689}}].

\bibitem{Ding:2009iy}
G.-J. Ding, {\it {Fermion Masses and Flavor Mixings in a Model with $S_4$
  Flavor Symmetry}},  {\em Nucl. Phys.} {\bf B827} (2010) 82--111,
  [\href{http://xxx.lanl.gov/abs/0909.2210}{{\tt arXiv:0909.2210}}].

\bibitem{Meloni:2009cz}
D.~Meloni, {\it {A See-Saw $S_4$ model for fermion masses and mixings}},  {\em
  J. Phys.} {\bf G37} (2010) 055201,
  [\href{http://xxx.lanl.gov/abs/0911.3591}{{\tt arXiv:0911.3591}}].

\bibitem{Morisi:2010rk}
S.~Morisi and E.~Peinado, {\it {An S4 model for quarks and leptons with maximal
  atmospheric angle}},  {\em Phys. Rev.} {\bf D81} (2010) 085015,
  [\href{http://xxx.lanl.gov/abs/1001.2265}{{\tt arXiv:1001.2265}}].

\bibitem{Dutta:2009bj}
B.~Dutta, Y.~Mimura, and R.~N. Mohapatra, {\it {An SO(10) Grand Unified Theory
  of Flavor}},  {\em JHEP} {\bf 05} (2010) 034,
  [\href{http://xxx.lanl.gov/abs/0911.2242}{{\tt arXiv:0911.2242}}].

\bibitem{Gatto:1968ss}
R.~Gatto, G.~Sartori, and M.~Tonin, {\it Weak selfmasses, {C}abibbo angle, and
  broken {SU$(2)\times$ SU$(2)$}},  {\em Phys. Lett.} {\bf B28} (1968)
  128--130.

\bibitem{Oakes:1969vm}
R.~J. Oakes, {\it {SU(2) x SU(2) breaking and the Cabibbo angle}},  {\em Phys.
  Lett.} {\bf B29} (1969) 683--685.

\bibitem{Ponton:2001hq}
E.~Ponton and E.~Poppitz, {\it {Casimir energy and radius stabilization in five
  and six dimensional orbifolds}},  {\em JHEP} {\bf 06} (2001) 019,
  [\href{http://xxx.lanl.gov/abs/hep-ph/0105021}{{\tt hep-ph/0105021}}].

\bibitem{Buchmuller:2009er}
W.~Buchmuller, R.~Catena, and K.~Schmidt-Hoberg, {\it {Enhanced Symmetries of
  Orbifolds from Moduli Stabilization}},  {\em Nucl. Phys.} {\bf B821} (2009)
  1--20, [\href{http://xxx.lanl.gov/abs/0902.4512}{{\tt arXiv:0902.4512}}].

\bibitem{Asaka:2002nd}
T.~Asaka, W.~Buchmuller, and L.~Covi, {\it {Exceptional coset spaces and
  unification in six dimensions}},  {\em Phys. Lett.} {\bf B540} (2002)
  295--300, [\href{http://xxx.lanl.gov/abs/hep-ph/0204358}{{\tt
  hep-ph/0204358}}].

\bibitem{Hall:2001xr}
L.~J. Hall, Y.~Nomura, T.~Okui, and D.~Tucker-Smith, {\it {SO(10) unified
  theories in six dimensions}},  {\em Phys. Rev.} {\bf D65} (2002) 035008,
  [\href{http://xxx.lanl.gov/abs/hep-ph/0108071}{{\tt hep-ph/0108071}}].

\bibitem{ArkaniHamed:2001tb}
N.~Arkani-Hamed, T.~Gregoire, and J.~G. Wacker, {\it {Higher dimensional
  supersymmetry in 4D superspace}},  {\em JHEP} {\bf 03} (2002) 055,
  [\href{http://xxx.lanl.gov/abs/hep-th/0101233}{{\tt hep-th/0101233}}].

\bibitem{Hebecker:2001jb}
A.~Hebecker and J.~March-Russell, {\it {The structure of GUT breaking by
  orbifolding}},  {\em Nucl. Phys.} {\bf B625} (2002) 128--150,
  [\href{http://xxx.lanl.gov/abs/hep-ph/0107039}{{\tt hep-ph/0107039}}].

\bibitem{Asaka:2002my}
T.~Asaka, W.~Buchmuller, and L.~Covi, {\it {Bulk and brane anomalies in six
  dimensions}},  {\em Nucl. Phys.} {\bf B648} (2003) 231--253,
  [\href{http://xxx.lanl.gov/abs/hep-ph/0209144}{{\tt hep-ph/0209144}}].

\bibitem{Mirabelli:1997aj}
E.~A. Mirabelli and M.~E. Peskin, {\it {Transmission of supersymmetry breaking
  from a 4- dimensional boundary}},  {\em Phys. Rev.} {\bf D58} (1998) 065002,
  [\href{http://xxx.lanl.gov/abs/hep-th/9712214}{{\tt hep-th/9712214}}].

\bibitem{Asaka:2001eh}
T.~Asaka, W.~Buchmuller, and L.~Covi, {\it {Gauge unification in six
  dimensions}},  {\em Phys. Lett.} {\bf B523} (2001) 199--204,
  [\href{http://xxx.lanl.gov/abs/hep-ph/0108021}{{\tt hep-ph/0108021}}].

\bibitem{Bazzocchi:2009pv}
F.~Bazzocchi, L.~Merlo, and S.~Morisi, {\it {Fermion Masses and Mixings in a
  S4-based Model}},  {\em Nucl. Phys.} {\bf B816} (2009) 204--226,
  [\href{http://xxx.lanl.gov/abs/0901.2086}{{\tt arXiv:0901.2086}}].

\bibitem{Dienes:1998vh}
K.~R. Dienes, E.~Dudas, and T.~Gherghetta, {\it {Extra spacetime dimensions and
  unification}},  {\em Phys. Lett.} {\bf B436} (1998) 55--65,
  [\href{http://xxx.lanl.gov/abs/hep-ph/9803466}{{\tt hep-ph/9803466}}].

\bibitem{Dienes:1998vg}
K.~R. Dienes, E.~Dudas, and T.~Gherghetta, {\it {Grand unification at
  intermediate mass scales through extra dimensions}},  {\em Nucl. Phys.} {\bf
  B537} (1999) 47--108, [\href{http://xxx.lanl.gov/abs/hep-ph/9806292}{{\tt
  hep-ph/9806292}}].

\bibitem{Calibbi:2009cp}
L.~Calibbi, L.~Ferretti, A.~Romanino, and R.~Ziegler, {\it {Gauge coupling
  unification, the GUT scale, and magic fields}},  {\em Phys. Lett.} {\bf B672}
  (2009) 152--157, [\href{http://xxx.lanl.gov/abs/0812.0342}{{\tt
  arXiv:0812.0342}}].

\bibitem{Georgi:1979df}
H.~Georgi and C.~Jarlskog, {\it {A New Lepton - Quark Mass Relation in a
  Unified Theory}},  {\em Phys. Lett.} {\bf B86} (1979) 297--300.

\bibitem{Dutta:2009ij}
B.~Dutta, Y.~Mimura, and R.~N. Mohapatra, {\it {Origin of Quark-Lepton Flavor
  in SO(10) with Type II Seesaw}},  {\em Phys. Rev.} {\bf D80} (2009) 095021,
  [\href{http://xxx.lanl.gov/abs/0910.1043}{{\tt arXiv:0910.1043}}].

\bibitem{King:2005bj}
S.~F. King, {\it {Predicting neutrino parameters from SO(3) family symmetry and
  quark-lepton unification}},  {\em JHEP} {\bf 08} (2005) 105,
  [\href{http://xxx.lanl.gov/abs/hep-ph/0506297}{{\tt hep-ph/0506297}}].

\bibitem{Masina:2005hf}
I.~Masina, {\it {A maximal atmospheric mixing from a maximal CP violating
  phase}},  {\em Phys. Lett.} {\bf B633} (2006) 134--140,
  [\href{http://xxx.lanl.gov/abs/hep-ph/0508031}{{\tt hep-ph/0508031}}].

\bibitem{Antusch:2005kw}
S.~Antusch and S.~F. King, {\it {Charged lepton corrections to neutrino mixing
  angles and CP phases revisited}},  {\em Phys. Lett.} {\bf B631} (2005)
  42--47, [\href{http://xxx.lanl.gov/abs/hep-ph/0508044}{{\tt
  hep-ph/0508044}}].

\bibitem{Antusch:2003kp}
S.~Antusch, J.~Kersten, M.~Lindner, and M.~Ratz, {\it Running neutrino masses,
  mixings and {CP} phases: Analytical results and phenomenological
  consequences},  {\em Nucl. Phys.} {\bf B674} (2003) 401--433,
  [\href{http://xxx.lanl.gov/abs/hep-ph/0305273}{{\tt hep-ph/0305273}}].

\end{thebibliography}\endgroup


\end{document}